\documentclass[letter,center,fleqn]{NAR}

\usepackage[caption=false]{subfig} %multi figures
\usepackage{amssymb} % Adds new symbols to be used in math mode
\usepackage{mathtools}
\usepackage{color}

\newcommand{\dGmeasure}{\Delta\Delta G}
\newcommand{\Dout}{D_{out}}

\copyrightyear{}
\pubdate{}
\begin{document}

\title{RNA structure generates natural cooperativity between single-stranded RNA binding proteins targeting 5' and 3'UTRs}

\author{%
Yi-Hsuan Lin\,$^{1}$ and
Ralf Bundschuh\,$^{1,2,3,4}$
%and Second Co-Author\,$^2$%
\footnote{To whom correspondence should be addressed. Tel: +1 (614) 688 3978; Fax: +1 (614) 292 7557; Email: bundschuh@mps.ohio-state.edu}
}

\address{%
$^{1}$Department of Physics, 
$^{2}$Department of Chemistry \& Biochemistry,
$^{3}$Division of Hematology,
and
$^{4}$Center for RNA Biology,
The Ohio State University, 191 W Woodruff Ave., 
Columbus, OH 43210-1107, USA
}
% Affiliation must include:
% Department name, institution name, full road and district address,
% state, Zip or postal code, country

%\history{%
%Received January 1, 2009;
%Revised February 1, 2009;
%Accepted March 1, 2009}

\maketitle

\begin{abstract}
In post-transcriptional regulation, an mRNA molecule is bound by many proteins and/or miRNAs to modulate its function. To enable combinatorial gene regulation, these binding partners of an RNA must communicate with each other, exhibiting cooperativity. Even in the absence of direct physical interactions between the binding partners, such cooperativity can be mediated through RNA secondary structures, since they affect the accessibility of the binding sites. Here we propose a quantitative measure of this structure-mediated cooperativity that can be numerically calculated for an arbitrary RNA sequence. Focusing on an RNA with two binding sites, we derive a characteristic difference of free energy differences, i.e. $\Delta\Delta G$, as a measure of the effect of the occupancy of one binding site on the binding strength of another. We apply this measure to a large number of human and C. elegans mRNAs, and find that structure-mediated cooperativity is a generic feature. Interestingly, this cooperativity not only affects binding sites in close proximity along the sequence but also configurations in which one binding site is located in the 5'UTR and the other is located in the 3'UTR of the mRNA. Furthermore, we find that this end-to-end cooperativity is determined by the UTR sequences while the sequences of the coding regions are irrelevant.

\end{abstract}

\section{Introduction}

Post-transcriptional regulation of messenger RNAs (mRNAs) is an
important component in cellular information processing.  At the
molecular level, this post-transcriptional regulation involves the
binding of a multitude of proteins and other factors such as microRNAs
incorporated into riboprotein complexes to a messenger RNA
molecule~\cite{Burd94, Glisovic08}.  These proteins affect gene
expression through regulation of mRNA stability, localization, and
translation. The importance of these RNA protein interactions can be
gauged from the fact that there are  
%approximately 400 RNA binding proteins annotated in the human proteome~\cite{Cook10}
close to 800 RNA binding proteins annotated in the proteome of the human embryonic kidney~\cite{Baltz12} and 860 in HeLa cells~\cite{Castello12},
and that a PubMed title search for the words ``RNA'', ``protein'', and
``binding'' yields more than $2{,}500$ articles.

In order to obtain {\em combinatorial} gene regulation, binding events of
RNA binding proteins and other RNA binding factors have to be
interdependent, or in biochemical terms {\em cooperative}.  Since most
protein and microRNA binding sites are located in the untranslated
regions (UTRs) at the ends of a messenger RNA~\cite{Mignone02}, it is
an interesting question how cooperativity between these distant
regions of an RNA molecule is achieved, especially in light of recent
evidence that the protein bridge between the 3' and 5'UTRs in
eukaryotes mediated by proteins that simultaneously bind the poly(A)
tail and the cap may not be essential~\cite{Park11}.  Here, we present
a natural mechanism for such cooperativity that relies on RNA
secondary structure alone.

\begin{figure}
\includegraphics[width=0.95\columnwidth]{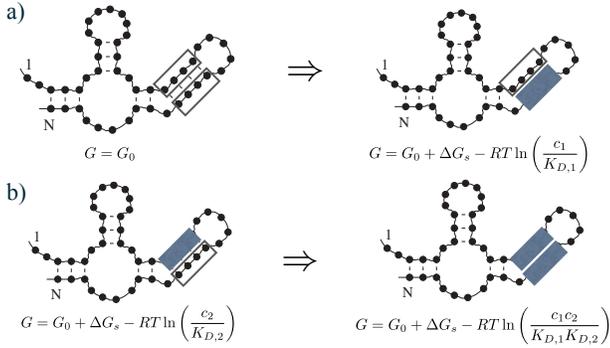}
\caption{An example of structure-mediated cooperativity, where open
  boxes indicate unoccupied protein binding sites and filled boxes
  indicate occupied protein binding sites.  $G$ is the free energy of
  the RNA-protein complex in each step, $G_0$ is the free energy of
  the RNA secondary structure alone, $\Delta G_s$ is the free energy
  associated with breaking the base pairs in the stem containing the
  two protein binding sites, $c_1$, $c_2$, $K_{D,1}$, and $K_{D,2}$
  are the concentrations and dissociation constants of the first and
  second protein, repectively, $R$ is the gas constant, and $T$ is the
  temperature in Kelvin. (a) To have the first protein bind to bases
  in a stem of the structure in the absence of the second protein,
  the stem has to be broken, resulting in a free energy difference of
  $\Delta G_{0\to1} = \Delta G_s-RT\ln(c_1/K_{D,1})$. (b) Once the
  second protein is bound to the opposing bases in the stem, the first
  protein can directly bind to its binding site without the loss of
  free energy by breaking additional base pairs, yielding $\Delta G_{2\to12} =
  -RT\ln(c_1/K_{D,1}) < \Delta G_{0\to1}$. Thus, in this case the two
  proteins exhibit positive cooperativity.}
	\label{fig:car_coop}
\end{figure}

Qualitatively the mechanism for this cooperativity is as follows: Many
RNA binding proteins bind to single-stranded regions of
RNA~\cite{Antson00, Ray09, Ray13}.  Therefore, a natural competition arises between
the formation of intramolecular base pairs in the RNA and the binding
of proteins to unpaired bases (here and in the following we will
restrict our language to ``RNA binding proteins'' even though in the
context of this study any RNA binding factor that competes with the
RNA for base pairing, such as, e.g., a microRNA, can play the role of
what we call an ``RNA binding protein'').  We have recently discovered
that this competition between binding to a single stranded RNA binding
protein and formation of intramolecular base pairs provides a possible
mechanism for cooperativity among proteins binding an RNA
molecule~\cite{Lin13}.  A protein binding an RNA molecule will change
the RNA's secondary structure by prohibiting the bound bases from
pairing, which in turn affects the base pairing at the binding site of
the second protein, and thus the ability of the second protein to bind
(see Fig.~\ref{fig:car_coop}).  In our previous work~\cite{Lin13} we
have focused on establishing that this RNA secondary structure
mediated cooperativity is a long-range effect, and used simplified
models of RNA folding to understand this behavior for random
sequences.  Here, we will show that this cooperativity (i) occurs for
natural sequences, (ii) is of a biologically relevant order of
magnitude of several kcal/mol, (iii) not only occurs between
protein binding sites that are within 100nt or so along the RNA but
also between the 5'UTR and the 3'UTR, and (iv) is determined by the UTR sequences with little effect from the coding regions.

% **************************************************************
% Keep this command to avoid text of first page running into the
% first page footnotes
\enlargethispage{-65.1pt}
% **************************************************************

\section{MATERIALS AND METHODS}

\subsection{RNA secondary structure prediction}

A single-stranded RNA folds into different configurations by pairing
its nucleotides and forming stable Watson-Crick (i.e. A--U or G--C) or
wobble (i.e., G--U) base pairs. Stacks of these base pairs form stiff
helices connected by flexible linkers of unpaired bases.  In the case
of structural RNAs these elements fold into a three-dimensional
structure required for the biological function of the molecule, called
the tertiary structure.  However, given the generic propensity of
nucleotides to base pair, also messenger RNAs, which are not primarily
designed to fold into a specific structure, will form base pairs.
Since here we are interested in messenger RNAs, we will focus on these
secondary structures only.  For computational convenience we follow
the usual approach of the field and exclude the formation of
pseudoknots because such structures contribute small amounts of free
energy if they are short and become kinetically suppressed if they are
long~\cite{Higgs00}.  More specifically, we use the Vienna
package~\cite{Hofacker94} to calculate the free energies of secondary
structure folding, which implements the state of the art nearest
neighbor free energy model of RNA secondary structure
formation~\cite{Mathews99}.

\subsection{RNA-protein binding}

To discuss RNA-protein binding events, we consider only the simplest effects of the binding of the protein on the RNA secondary structure~\cite{Lin13,Forties10}: (i) A bound protein prevents all nucleotides in its footprint from base pairing and (ii) binding of a protein to the RNA results in a free energy gain of $RT\ln(c/K_D)$, where $c$ is the concentration of the protein in solution, $K_D$ is the dissociation constant of the protein from the specific site in the RNA, which contains all the details of the interaction between the protein and the RNA, $R$ is the gas constant and $T$ is the temperature in Kelvin~\cite{cite:RTlnc}.  More sophisticated effects of RNA-protein binding, for example the increased geometrical distance of the bases adjacent to the protein binding site upon binding of the protein or any other modifications to the binding propensities of nucleotides not within the footprint of the protein, are not considered in our minimal model and will be subject of future research.

\begin{figure}
\includegraphics[width=\columnwidth]{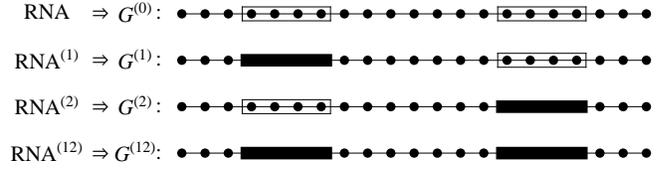}
\caption{The four possible RNA-protein complexes to consider. Lines are RNA backbones and dots are bases. Transparent and black blocks represent the unoccupied and occupied protein binding sites, respectively. Only the bases which are not in the occupied binding regions participate in base pairing.}
	\label{fig:Z0Z1Z2Z12}
\end{figure}

We will further constrain ourselves to the case that there are only two proteins, $P_1$ and $P_2$, binding one binding site each on a given RNA molecule, which is the simplest system to investigate protein-protein cooperativity in~\cite{Lin13}.  Consequently, there are four possible RNA-protein complexes, distinguished by whether the first and/or the second protein binding site is occupied or not, which we label as $RNA$, $RNA^{(1)}$, $RNA^{(2)}$, and $RNA^{(12)}$ as illustrated in Fig.~\ref{fig:Z0Z1Z2Z12}.  For each complex, different sets of base pairs can participate in base pairing interactions.  We calculate the free energies $G^{(0)}$, $G^{(1)}$, $G^{(2)}$, and $G^{(12)}$ of the secondary structures for each of the four complexes using the constraint folding capabilities of the Vienna package to exclude the nucleotides ``hidden'' in the footprints of bound proteins from base pairing.  To obtain the free energies of the entire complexes the protein binding free energy $RT\ln(c_1/K_{D,1})$ has to be subtracted from $G^{(1)}$ and $G^{(12)}$ and the protein binding free energy $RT\ln(c_2/K_{D,2})$ of the second protein has to be subtracted from $G^{(2)}$ and $G^{(12)}$.

\subsection{Quantitating protein-protein cooperativity}

With the two proteins $P_1$ and $P_2$, all RNA-protein binding reactions can be expressed as
\begin{eqnarray*}
\begin{array}{ccc}
RNA+P_1+P_2 & \xleftrightarrow{ \Delta G_{0\to1}
} & RNA^{(1)} + P_2  \\
\\
\scriptstyle \Delta G_{0\to2}
\left\updownarrow\rule{0cm}{1cm}\right.  \qquad & ~ & \qquad
\left\updownarrow\rule{0cm}{1cm}\right. 
\scriptstyle \Delta G_{1\to12} \\ 
\\
RNA^{(2)}+P_1 & \xleftrightarrow[ 
\scriptstyle \Delta G_{2\to12}]{} & \quad RNA^{(12)}  \\
\end{array},
\end{eqnarray*}
where the free energy differences $\Delta G$ of the four reactions are given by
\begin{subequations}
\begin{align}
& \Delta G_{0\to1} = G^{(1)}-G^{(0)}-RT\ln(c_1/K_{D,1}), \\
& \Delta G_{0\to2} = G^{(2)}-G^{(0)}-RT\ln(c_2/K_{D,2}), \\
& \Delta G_{1\to12} = G^{(12)}-G^{(1)}-RT\ln(c_2/K_{D,2}), \\
& \Delta G_{2\to12} = G^{(12)}-G^{(2)}-RT\ln(c_1/K_{D,1}).
\end{align}
\end{subequations}

The cooperativity between the two binding sites can be quantified by
comparing the binding free energies $\Delta G$ of the reactions on
opposite sides of this diagram, e.g., the difference in binding free
energy $\Delta G_{0\to1}$ for the binding of protein $P_1$ in the
absence of protein $P_2$ and the binding free energy $\Delta
G_{2\to12}$ associated with binding of protein $P_1$ in the presence
of protein $P_2$ (or vice versa with the two proteins interchanged
which leads to the same result).  This yields the cooperativity free
energy
\begin{equation}
\begin{aligned}
\dGmeasure & \equiv \Delta G_{0\to1} - \Delta G_{2\to12} = \Delta G_{0\to2} - \Delta G_{1\to12} \\
& = G^{(1)}+G^{(2)}-G^{(12)}-G^{(0)}
\end{aligned}
	\label{eq:dGmeasure}
\end{equation}
which becomes independent of the protein concentrations $c_1$ and $c_2$. Therefore, for an arbitrary RNA sequence, once the two protein binding sites are assigned, the Vienna package can be used to calculate the structural free energies $G^{(0)}$, $G^{(1)}$, $G^{(2)}$, and $G^{(12)}$, from which the free energy difference $\dGmeasure$ quantifying the cooperativity can be deduced via Eq.~(\ref{eq:dGmeasure}). For a positive cooperativity, a protein already bound helps the other protein bind, leading to $\Delta G_{2\to12} < \Delta G_{0\to1}$ and thus a positive $\dGmeasure$; for a negative cooperativity, the situation is reversed and $\dGmeasure$ is negative. 

\subsection{Locations of protein binding sites}

\begin{figure}
\includegraphics[width=0.8\columnwidth]{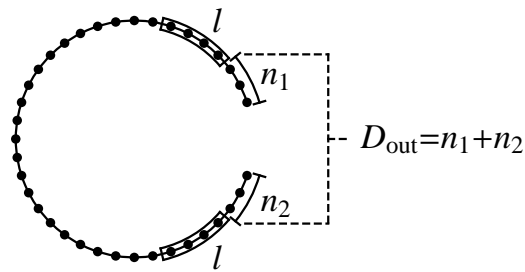}
\caption{Definition of the outside distance, $\Dout$, as the sum of the distances from each of the two protein binding sites to their corresponding ends in units of nt. Black dots represent bases and the two blocks represent the protein binding sites with footprint length $l$ (in this figure $l=4$nt). In our numerical calculation we restrict ourselves to the symmetric case $n_1 = n_2$, and use a footprint size of $l=$6nt.}
	\label{fig:Dout}
\end{figure}

We are especially interested in the cooperativity between the proteins
bound on the 5' and 3'UTRs close to the 5' and 3' ends of the
mRNAs, respectively. Therefore, we assign the two protein binding
sites of each mRNA in their 5' and 3' ends, and vary their distances
from the ends of the RNAs (note that in our minimal model the binding
site of each protein is specified explicitly, i.e., that we assume
perfect sequence specificity for the sequence at whatever binding site
we choose without actually specifying that sequence).  The relative
location of the two binding sites is quantified by the {\em outside
  distance}, $\Dout$, which is defined in Fig. \ref{fig:Dout} as the
sum $\Dout = n_1 + n_2$ of the distances $n_1$ and $n_2$ of the two
binding sites from their respective sequence ends.  In order to avoid
the very significant computational cost of varying the two distances
$n_1$ and $n_2$ independently, we consider only the situation that the
two binding sites are symmetric, $n_1 = n_2$, 
when we calculate the $\dGmeasure(\Dout)$ of an ensemble of mRNAs or of several extremely long sequences; we do not expect any
qualitative differences for asymmetric situations. In our
studies involving microRNA binding sites described below, we choose the known
microRNA binding site in the 3'UTR and vary the location of the putative protein
binding site in the 5'UTR. The footprint of
the proteins is set to be $l=$6nt in agreement with typical footprint
sizes of RNA binding proteins.

\subsection{Sequences} 

We select human mRNA sequences in NCBI's RefSeq database~\cite{NCBI} to study cooperativity between 5' and 3'UTRs. As a comparison to gauge the influence of the length of the UTRs, we also investigate {\it C. elegans} mRNAs from the WormBase database~\cite{WormBase}, in which the UTRs are much shorter than those in human mRNAs. 

Some of the sequences are deposited in these databases including their poly(A) tails.  Since these tails, as homopolymers, contribute very little to the RNA secondary structure, and in order to treat all mRNAs consistently no matter if they were submitted to the databases with or without poly(A) tails, we discard all poly(A) tails. Specifically, we remove any runs of three or more consecutive adenines at the 3' ends of the sequences in the database.  We are aware that in eliminating the poly(A) tails, we lose the ability of applying our results to the important class of poly(A) binding proteins (PABP)~\cite{Gorlach94, Kahvejian01, Kahvejian05}; however, since the poly(A) tails should not form secondary structures, PABPs are not expected to be subject to structure-mediated cooperativity, and thus are irrelevant in this study.

Since RNA secondary structure prediction scales as the third power of sequence length, computational resources constrain us to focus on the shorter mRNAs in the database.  Dictated by computational feasibility, we thus choose the range from 500 to 1500 bases (counted after dropping the poly(A) tails). For human sequences, we select those sequences in which both 5' and 3'UTRs are longer than 110 bases to ensure that all our binding sites are always in the UTRs even for the greatest outside distance, $\max(\Dout) = 200$nt, yielding a total of 2282 sequences. For {\it C. elegans} sequences, we select the sequences with short UTRs by constraining the UTR lengths between 30 and 150 bases, yielding a total of 1277 sequences.

Since we want to identify which, if any, regions of the human mRNAs might have evolved to generate cooperativity, we also generate several sets of artificial RNA sequences based on our set of human sequences: (a) We shuffle each of the 2282 selected human sequences, generating an ensemble of random sequences with equal dinucleotide frequencies~\cite{Altschul85}. (b) We shuffle the UTRs of each of the 2282 human sequences to generate an ensemble of random sequences with equal dinucleotide frequencies in the UTRs but with unchanged coding sequences (CDSs). (c) We select sequences that have UTR lengths in the range from 110nt to 400nt and replace their CDSs by the CDS of the single sequence NM004242 (300nt), rendering an ensemble of totally 1377 artificial sequences with a fixed CDS.

In order to show that our results are not limited to short mRNA sequences, we also selected four individual mRNA sequences with more than 2000 bases, namely PTP4A1, MARCKS, MYC, and PPP1R15B for verification purposes (it is computationally feasibly to study a few individual long mRNAs, but averaging over the entirety or even just a large number of these is not possible).

\subsection{Human microRNA binding sites}

Above, we describe symmetric protein binding sites, which are designed to have $n_1 = n_2$ in $\Dout = n_1 +n_2$. However, in addition to these ``assigned" binding sites, we also investigate ``natural" microRNA binding sites on 3'UTRs~\cite{Bartel04}, and specifically their cooperativity with proteins binding on the 5'UTR.

We fetch microRNA binding sites from microRNA.org~\cite{microRNAorg} using a strict mirSVR score cutoff of $-1$~\cite{Betel10}. Again, we select only the mRNA sequences with total lengths from 500nt to 1500nt and UTR lengths above 110nt for computational reasons. We limit the distance between the microRNA binding site and the 3' end of the mRNA to be less than 100 bases, i.e. $n_2 \leq 100$, because we want to focus on the cooperativity between microRNAs bound on the end and proteins bound at the beginning of the same mRNA sequence. The above-mentioned conditions yield 1173 microRNA-mRNA binding pairs, in which some of the mRNA sequences correspond to more than one microRNA binding site. We investigate each such microRNA-mRNA pair independently, i.e. we consider only one specific microRNA interacting with its bound mRNA at a time. The cooperativity is calculated between the microRNA and a protein binding site around the end of the 5'UTR. We set the protein footprint to be $l = 6$nt, and investigate all possible protein binding sites from exactly at the end ($n_1 = 0$nt) to 100 bases from the end of 5'UTR ($n_1 = 100$nt). For each microRNA-mRNA pair we find the 5'UTR protein binding site that has the greatest $\dGmeasure$ with respect to the microRNA, recording it as the strongest cooperativity of the microRNA-mRNA pair.

As a comparison, multiple random ensembles of microRNA-mRNA binding pairs are generated. In each random ensemble the microRNA binding sites, which are located by their $n_2$ on their ``naturally" bound mRNA, are randomly assigned to a different mRNA. The strongest cooperativities of the reassigned microRNA-mRNA pairs are recorded as a reference to compare the cooperativities of the natural microRNA-mRNA pairs to.

\subsection{Artificial evolution of highly cooperative sequences}

We use an evolutionary algorithm, in order to explore how high
structure-mediate cooperativities can become. We choose five natural sequences, namely NM\_006308, NM\_022645, NM\_001104548, NM\_001244390, and NM\_001272086,
as starting sequences. We calculate the maximum over all $|\dGmeasure|$ 
%{\tt Is it correct that I added the absolute value?} 
for all possible pairs of protein binding sites. Considering the massive computational resources required for a search through all possible protein binding site pairs (on the order of $N^3 \times N^2 \sim N^5$ in which $N$ is the length of sequence), we compromise on merely searching protein binding sites in the region of $5{\rm nt} \leq n_1, n_2 \leq 25{\rm nt}$, as we are more interested in the cooperativitiy between the ends of the two UTRs. In each evolutionary step, we randomly choose and mutate a nucleotide of the sequence, and recalculate the maximum of $|\dGmeasure|$ for all protein binding pairs. If the maximum is greater than that of the previous sequence, we keep the new sequence; otherwise we continue with the previous sequence. This evolutionary step is repeated $1,000$ times. %{\tt Is it 1000 steps or 1000 succesful mutations?}.

\section{RESULTS}

\subsection{Distance dependence of structure-mediated cooperativity}

We quantify RNA secondary structure mediated protein binding
cooperativity by the cooperativity free energy $\dGmeasure$, which is
defined as the difference between the free energies of binding of the
first protein in the presence and absence of the second protein (see
Materials and Methods).  If it is positive, the presence of the second
protein promotes binding of the first while if it is negative, the
presence of the second protein prevents binding of the first.  In
order to demonstrate that RNA secondary structure mediated protein
binding cooperativity is a generic effect providing cooperativity
between proteins binding at the two ends of an RNA molecule, we select
$2282$ human mRNA sequences (see Materials and Methods).  We choose
one protein binding site to be close to the 5' end of each molecule and one
close to the 3' end and systematically vary their distance from their
respective ends such that the sum $\Dout$ of the two distances from
the ends ranges from $0$nt to $200$nt (see Fig.~\ref{fig:Dout}).  We
calculate the cooperativity free energy $\dGmeasure$ for each
``outside'' distance $\Dout$ and each of the $2282$ sequences.

\begin{figure}
\includegraphics[width=\columnwidth]{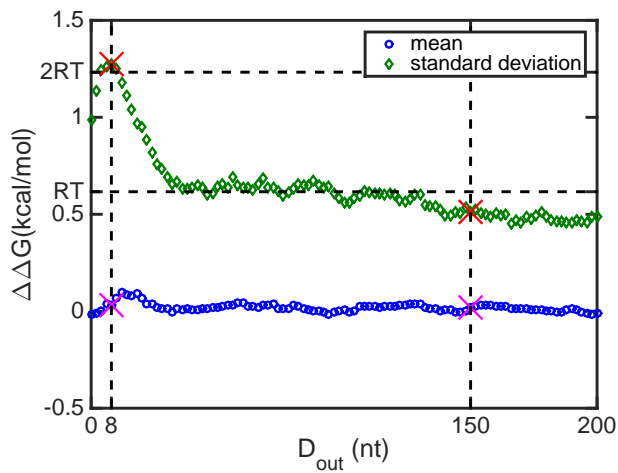}
\caption{The mean (circles) and standard deviation (diamonds) of the cooperativity free energy $\dGmeasure$ of 2282 human sequences, as a function of the outside distance $\Dout$ between the two protein binding sites.  The mean of the cooperativity free energy $\dGmeasure$ is close to zero for all outside distances, while its standard deviation is on the order of or greater than the thermal energy scale of $R T = $0.6kcal/mol.  The dashed vertical lines label two different outside distances $\Dout$ for which the whole distribution of the cooperativity free energies are compared in Fig.~\ref{fig:dGhistComp_2D}.  The maximum of the standard deviation of the cooperativity free energy occurs at $\Dout=$8nt.}
	\label{fig:dG_AVGSTD}
\end{figure}

Fig.~\ref{fig:dG_AVGSTD} shows the mean and standard deviation of the
cooperativity free energy $\dGmeasure$ over all $2282$ sequences as a
function of the outside distance $\Dout$ of the two protein binding
sites.  The mean cooperativity free energy remains within about
$0.1$kcal/mol of zero, more or less independent of the outside distance
between the protein binding sites.  That in principle leaves two
possibilities, namely that either RNA secondary structure mediated
cooperativity is a very small effect or that it is a significant
effect but roughly equal numbers of sequences show positive as
negative cooperativity.  The standard deviation allows us to
differentiate between these two scenarios: the fact that it starts out
above 1kcal/mol when the two binding sites are very close to their
respective ends and then falls to about 0.5kcal/mol for outside
distances of $30$nt and stays in this range for binding sites up to
100nt away from their respective ends (yielding a total outside distance
of 200nt) implies that many sequences must have cooperativity free
energies with absolute values of this order of magnitude.

\subsection{Structure-mediated cooperativity is significant for
large numbers of mRNA molecules}

\begin{figure}
\includegraphics[width=\columnwidth]{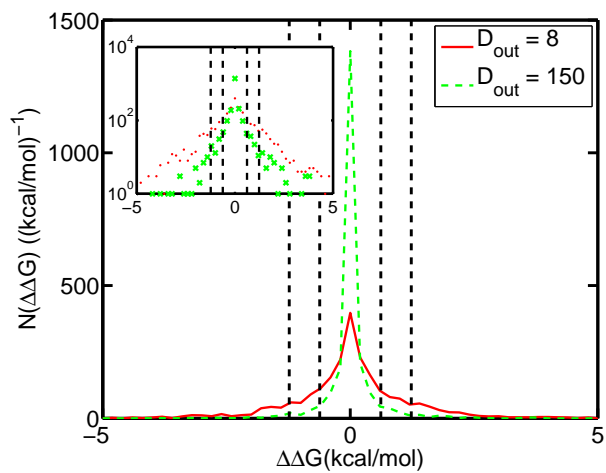}
\caption{The number of mRNA molecules with respect to given cooperativity free energies $\dGmeasure$ (in bin sizes of 0.2kcal/mol) at outside distances between the protein binding sites of $\Dout=$8 (red solid line) and $\Dout=$150 (green dashed line). The inset shows the same data on a logarithmic scale. The two distributions are both centered at $0$.  The distribution for $\Dout=$8 (when the protein binding sites are close to their respective ends of the molecule) extends much wider than that for $\Dout=$150 when the protein binding sites are further away from the ends of the molecule.  Vertical black dashed and dash-dotted lines label $|\dGmeasure|=RT$ and $|\dGmeasure|=2RT$, respectively. }
	\label{fig:dGhistComp_2D}
\end{figure}

In order to classify cooperativity free energies as significant or
not, it is convenient to compare them to the thermal energy of $RT$,
which is approximately 0.6kcal/mol at temperatures of $37^{\circ}$C.
Free energy differences below this magnitude will not have a
measurable effect on protein binding in the thermal environment of a
cell.  Free energy differences larger than $RT$ do result in
significant protein binding cooperativity.  The horizontal dashed
lines in Fig.~\ref{fig:dG_AVGSTD} indicate cooperativity free energy
levels of $RT$ and $2RT$ and show that the standard deviation of the
cooperativity free energy is on the order of $RT$ in the distance
range of $30{\rm nt} \lesssim\Dout\lesssim150{\rm nt}$.  Thus a good fraction of
molecules must have cooperativity free energies larger than $RT$ even
at these outside distances.  This is illustrated in
Fig.~\ref{fig:dGhistComp_2D} for two specific outside distances, namely
$\Dout=8$nt where both proteins are very close to their respective
ends and the standard deviation of the cooperativity free energy has
its maximum, and $\Dout=150$nt, where the standard deviation of the
cooperativity free energy starts to fall below $RT$.  The figure
illustrates how many sequences have a given cooperativity free energy
in bin sizes of 0.2kcal/mol. 
%\texttt{The bin size can be removed if the y axis of the figure is actually normalized such that the integral is 5630)}. 
As suggested by the small mean, the
distributions of cooperativity free energies are symmetric with equal
numbers of molecules showing positive as negative cooperativity.  It
can also be seen that the distribution at $\Dout=8$nt is much wider
than the distribution at $\Dout=150$nt as expected from their standard
deviations shown in Fig.~\ref{fig:dG_AVGSTD}.  The inset of
Fig.~\ref{fig:dGhistComp_2D} shows the same distributions as the main
figure but on a logarithmic scale that emphasizes the long tails of
the distribution indicating some sequences with very large
cooperativity free energies.

\begin{figure}
\includegraphics[width=\columnwidth]{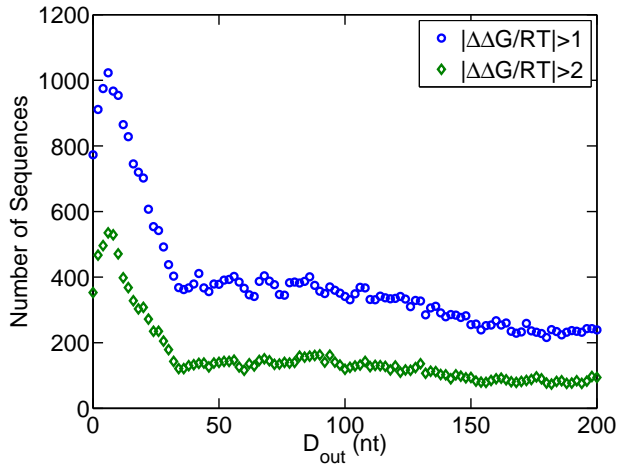}
\caption{The number of mRNA sequences with biochemically relevant protein binding cooperativity among the 2282 investigated human mRNA sequences. The two data sets correspond to $|\dGmeasure|/k_B T > 1$ (circles) and $|\dGmeasure|/k_B T > 2$ (diamonds).  The data shows that at any given distance $\Dout$ a significant fraction of molecules has a significant cooperativity free energy.}
	\label{fig:NofSeq}
\end{figure}

An additional way to quantify the significance of the cooperativity
free energies is to count the number of molecules out of the $2282$
tested that have a cooperativity free energy above a certain threshold
for each outside distance $\Dout$.  Fig.~\ref{fig:NofSeq} shows
this data for thresholds of $RT$ and $2RT$.  The data shows that when
the proteins bind very close to their respective ends of the RNA
molecule, about half of the totally 2282 investigated sequences have
$|\dGmeasure| > RT$ and a quarter have $|\dGmeasure| > 2RT$ (at
$\Dout$=6nt), implying that a significant cooperativity between two
protein binding sites, of which one is at the 5' end of a molecule and
the other is at the 3' end of a molecule, is a generic property among
messenger RNAs.  In addition, even at an outside distance of 200nt (each
protein binds 100nt from its respective end of the RNA) still 10\% of
the 2282 messenger RNAs studied show a cooperativity free energy of at
least $RT$ and more than a hundred show a cooperativity of at least
$2RT$.

\subsection{Cooperativities in human mRNAs and random RNA sequences have different statistical characteristics}

In our previous work we verified that structure mediated cooperativity is a generic property of all RNA molecules, stating that cooperativity exists in an ensemble of random RNA sequences~\cite{Lin13}. In this study we investigate the cooperativity in human, and thus evolved, mRNAs. Thus, we are curious about what the similarities and differences between the cooperativities for a ``random'' and a ``natural" sequence ensemble are.

%  Moreover, since we are interested in the cooperativity between proteins on UTRs, we also compare to the cooperativity in sequence ensembles in which the coding sequences (CDSs) are natural while the UTRs have been replaced by random sequences, or in which the CDS in each natural mRNA is replaced by a fixed CDS. All these investigations aim at clarifying the relationship between the random and natural sequences and the roles of UTRs and CDSs in the occurrence of the cooperativity.

To compare with the human mRNA sequences, we thus randomly permute nucleotides in each of the 2282 selected human sequences, and generate two ensembles of 2282 random sequences (see Materials and Methods): (a) random sequences generated by shuffling the entire sequence of each human mRNA such that the dinucleotide frequencies are identical to those of the original human mRNAs, and (b) random sequences generated by shuffling only the 5' and 3'UTR sequences of each human mRNA while leaving the CDS unaffected. We investigate cooperativity in these two random sequence ensembles, and compare their statistical properties with those of the natural human sequences.

\begin{figure}
\includegraphics[width=\columnwidth]{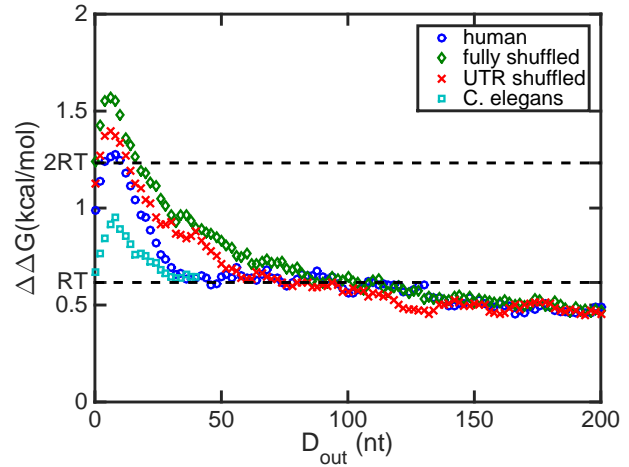}
\caption{The standard deviations of the cooperativity free energy $\dGmeasure$ for different mRNA ensembles including 2282 sequences each, as functions of the distance $\Dout$ between the protein binding sites.  Statistics of four sequence ensembles are shown: natural human mRNAs (circles), dinucleotide-shuffled human sequences (diamonds), sequences with randomized UTRs (crosses), and natural {\it C. elegans} mRNAs (squares). All four sequence ensembles have standard deviations on the order of or greater than the thermal energy scale of $R T = $0.6kcal/mol, implying that a biologically relevant $\dGmeasure$ is a general property for mRNA sequence ensembles. However, different sequence ensembles show different statistical characteristics.}
	\label{fig:STDcompare}
\end{figure}

The standard deviations of $\dGmeasure(\Dout)$ of the two random ensembles and the human mRNA ensemble are shown in Fig.~\ref{fig:STDcompare}. We notice that all these three ensembles comprise many sequences with biologically relevant cooperativities, i.e. $\dGmeasure$ is on the order of or greater than 0.6kcal/mol. However, the standard deviation of the human mRNA ensemble has a large peak at small $\Dout$, and then sharply decreases to merely around $RT$ at $\Dout \geq 50$nt. Instead, the standard deviation of the completely shuffled random sequences smoothly decays in the entire investigated region, $0{\rm nt} \leq \Dout \leq 200$nt. The standard deviation of the UTR-shuffled sequences is in between those of the human mRNA and fully shuffled sequence ensembles. 

\begin{figure}
\includegraphics[width=\columnwidth]{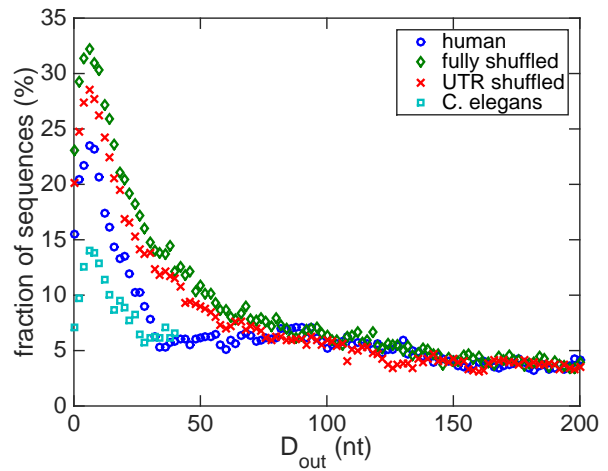}
\caption{The fractions of sequences that have $|\dGmeasure(\Dout)| > 2RT$ in different sequence ensembles: natural human mRNAs (circles), dinucleotide-shuffled human sequences (diamonds), sequences with randomized UTRs (crosses), and natural {\it C. elegans} mRNAs (squares). All four ensembles show maxima of the fraction at small $\Dout$. However, their distance dependences are different.}  %{\tt Change legends to ``human'', ``fully shuffled'', ``UTR shuffled'', and ``C. elegans''}}
	\label{fig:Nseq_2kT_compare}
\end{figure}

The fraction of sequences that have $| \dGmeasure | > 2RT$ in the human mRNA and the two random sequence ensembles are shown in Fig.~\ref{fig:Nseq_2kT_compare}. Similar to the standard deviations, for the two random sequence ensembles, the fractions of sequences with strong cooperativities decrease smoothly as $\Dout$ increases, whereas for human mRNAs the fraction decreases sharply once $\Dout$ becomes greater than that for the maximum at $\Dout = 8$nt. Taken together, a human mRNA is likely to have strong cooperativity between the ends of the 5' and 3'UTR, while this cooperativity decreases rapidly as the two UTR binding sites are farther from their respective ends. On the other hand, random sequences perform power-law like, smoothly decaying cooperativity with respect to $\Dout$.  We thus conclude that human RNAs are different from random RNAs in terms of their cooperativity properties.  While one might have expected stronger cooperativity in the evolved human sequences than in their randomly shuffled counterparts, we would like to point out that we, as discussed above, for computational reasons only look at symmetrically located binding sites of the two proteins.  These will in general not correspond to the true protein binding sites on those natural mRNAs and it is conceivable that a possible design for cooperativity at true binding sites in fact leads to the suppression of cooperativity at the symmetrically located positions we interrogate.

\subsection{Cooperativity is driven by the UTRs and not by the coding sequences}

\begin{figure}
\includegraphics[width=\columnwidth]{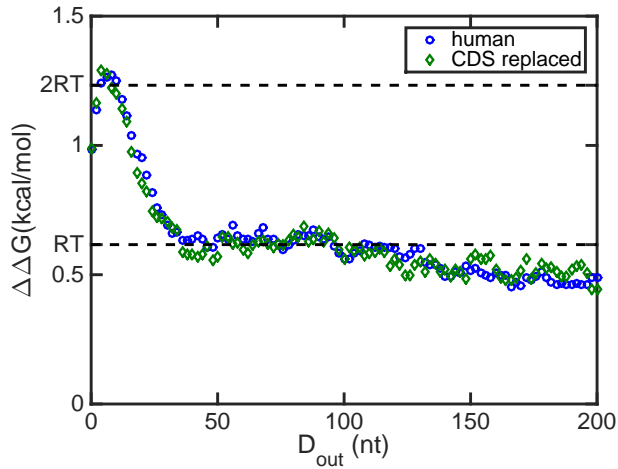}
\caption{The standard deviations of the cooperativity free energy $\dGmeasure$ of 2282 human mRNAs and 1377 CDS-replaced sequences. The two standard deviations are very similar to each other, implying that the cooperativity depends on UTRs instead of CDSs.}% {\tt change first label to ``human''.}}
	\label{fig:STDcompare_CDSreplace}
\end{figure}

Since we have shown in the previous section that replacing the UTRs with shuffled sequences yields similar cooperativities as shuffling the entire sequence, we here ask the inverse question and retain the UTRs while replacing the CDSs of the ensemble of human mRNAs by one fixed CDS (see Materials and Methods). Fig.~\ref{fig:STDcompare_CDSreplace} shows the comparison of the standard deviations of $\dGmeasure(\Dout)$ of human mRNAs and the CDS-replaced sequences. The two ensembles of sequences show very similar profiles. This similarity suggests that the cooperativity between the 5' and 3'UTR binding sites is determined by the two UTR sequences, while the coding region has little effect.

%We clarify the source of the cooperativity by replacing the CDSs of an ensemble of human mRNAs by a fixed CDS. To save the requirement of computational resources, we choose NM 004242 for supplying the fixed CDS because its coding region is relatively short, comprising only 300 bases. We select sequences of which the two UTRs are in the range from 110 to 400 bases among the 2282 investigated human mRNAs, and connect their UTRs to the CDS of NM 004242, rendering a total of 1377 sequences.

\subsection{Cooperativities in {\it C. elegans} mRNAs are weaker than those in human mRNAs}

Compared to human mRNA sequences, {\it C. elegans} mRNAs comprise much shorter UTRs~\cite{NCBI, WormBase}. As we have confirmed in the last section that the cooperativity is driven by the UTRs, it is natural to investigate the cooperativites of an ensemble of sequences that have short UTRs. Thus, we choose an ensemble of 1277 {\it C. elegans} mRNAs of which both UTRs are in the range from 30 to 150 bases and the total sequence lengths are between 500 and 1500 bases. Their standard deviation of $\dGmeasure(\Dout)$ and the number of sequences having significant $\dGmeasure(\Dout)$ are plotted in Figs.~\ref{fig:STDcompare} and \ref{fig:Nseq_2kT_compare}, respectively. The two figures show two consistent consequences. First, the cooperativities in {\it C. elegans} mRNAs are also biologically relevant, i.e. $|\dGmeasure| \gtrsim RT$. Second, although cooperativities in {\it C. elegans} mRNAs are biologically relevant, they are significantly weaker than the cooperativities in human sequences. The latter result is additional evidence for the strong relationship between cooperativity and UTR sequences.

\subsection{RNA structure mediated cooperativity between the
sequence ends persists in long mRNA molecules}

\begin{figure}
\includegraphics[width=\columnwidth]{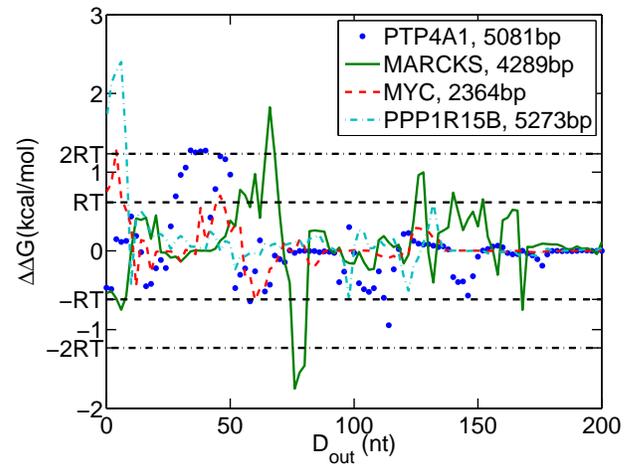}
\caption{The cooperativity free energy of four randomly selected mRNA molecules longer than 2000nt as a function of outside distance $\dGmeasure(\Dout)$.  Horizontal lines label the scales of $\pm RT$ and $\pm2RT$.  For all four molecule $\dGmeasure$ exceeds these energy scales and thus develops significant cooperativity at specific values of the outside distance $\Dout$. }
	\label{fig:dGplotIndptSeq}
\end{figure}

While computational complexity does not allow us to include mRNA
molecules longer than 1500nt in our systematic study above, we
selected four individual mRNA molecules with more than 2000nt to
determine if our findings of significant cooperativity between 5' and
3' ends of an mRNA hold for longer molecules as well.
Fig.~\ref{fig:dGplotIndptSeq} shows the cooperativity free energy of
these four molecules as a function of outside distance $\Dout$.  It
can be seen that all four chosen molecules have some outside distances
$\Dout$ at which $|\dGmeasure| > 2RT$ and even though for two of them
(MYC and PPP1R15B) the binding sites with such large cooperativity
free energy are very close to the ends, all molecules also have
cooperativity free energies in excess of $1RT$ at some outside
distance between $50$nt and $150$nt.

\subsection{Human microRNA binding sites show significant negative cooperativities}

A large number of microRNAs bind to the 3'UTRs of mRNA sequences to regulate their abundance. At the same time, proteins are bound to the 5'UTRs, e.g., see~\cite{Duncan06, Jungkamp11}. Thus, in addition to the symmetric binding sites with $n_1 = n_2$, studied so far, we also investigate the cooperativites between these microRNAs binding at their experimentally verified sites in the 3'UTR and putative 5'UTR binding proteins. We select 1173 microRNA-mRNA binding pairs from the microRNA.org database (see Materials and Methods). For each microRNA binding site in a 3'UTR, we search through putative protein binding sites in the first 100 bases in the 5'UTR of the corresponding mRNA, and record the cooperativity free energy $\dGmeasure$ of the binding site with the greastest absolute value of the cooperativity free energy $\dGmeasure$, termed $\dGmeasure_\mathrm{max}$. We note that although the position of the putative protein binding site in the 5'UTR is selected by maximizing the absolute value of the cooperativity free energy, the chosen cooperativity free energy still can be either positive or negative. As a comparison, we randomly shuffle the positions of the microRNA binding sites, matching them with different mRNAs. The $\dGmeasure_\mathrm{max}$ calculated from these randomly shuffled microRNA-mRNA binding pairs help us evaluate the significance of the characteristics discovered in the natural microRNA-mRNA pairs.

\begin{figure}
\includegraphics[width=\columnwidth]{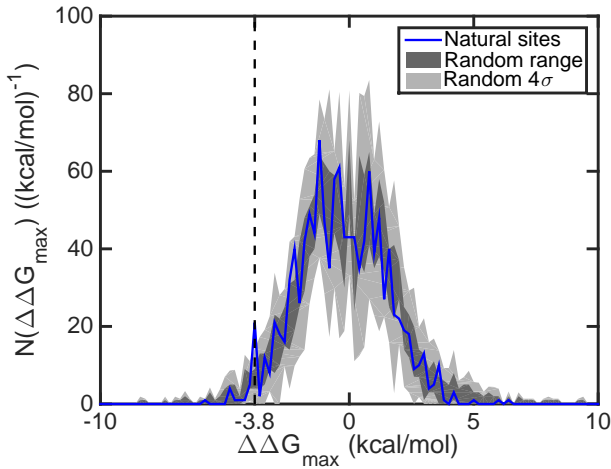}
\caption{The number of microRNA-mRNA binding pairs, for which the strongest cooperativity with respect to a binding site in the 5'UTR corresponds to a given value of $\dGmeasure_\mathrm{max}$, in bins of $0.2$kcal/mol. While the number distribution of human binding pairs does not show significant differences compared to randomly shuffled binding pairs for most values of $\dGmeasure_\mathrm{max}$, a peak at $\dGmeasure_\mathrm{max} \approx -3.8$kcal/mol suggest that there are a group of human sequences with strongly negative cooperativities between the 3'UTR-bound microRNA and putative 5'UTR-bound proteins.} %{\tt change max() to ${}_{max}$ on both axes. Put natural sites first in legend and spell correctly. Change STD to $\sigma$. Add ``range'' to dark gray legend.}}
	\label{fig:miRNA-mRNA_dGmax}
\end{figure}

Fig.~\ref{fig:miRNA-mRNA_dGmax} shows the distribution of the $\dGmeasure_\mathrm{max}$ in the 1173 microRNA-mRNA binding pairs in bin sizes of 0.2kcal/mol, with the grey regions representing the distribution of 10 ensembles of randomly shuffled binding pairs. The dark grey region for each cooperativity free energy bin designates the range of frequencies observed among the 10 ensembles while the light grey region indicates the regime of four standard deviations.  The figure illustrates that for most cooperativity free energies, the ensemble of natural binding pairs does not show significant differences from the ensembles of randomly shuffled binding pairs. However, in the region of negative $\dGmeasure_\mathrm{max}$, the natural binding pairs exhibit a characteristic difference from random fluctuations, namely a peak at $\dGmeasure_\mathrm{max} \approx -3.8{\rm kcal/mol} \approx -6RT$ labeled in Fig.~\ref{fig:miRNA-mRNA_dGmax}. The number of microRNAs with cooperativities of this magnitude is greater than its expected frequency based on the 10 random sequence ensembles by more than four standard deviations. We conclude that this peak is statistically significant, suggesting that a group of microRNAs result in strongly negative cooperativites with respect to proteins bound to the 5'UTRs. Considering the role of microRNAs in silencing and suppression of gene regulation~\cite{Bartel04}, the negative cooperativity we discover in the statistics of human microRNA-mRNA binding pairs is reasonable and, furthermore, implies that a microRNA can participate in combinatorial regulation with proteins binding far from its binding site.

\subsection{Cooperativities of artifically evolved RNA sequences can be large}

\begin{figure}
\includegraphics[width=\columnwidth]{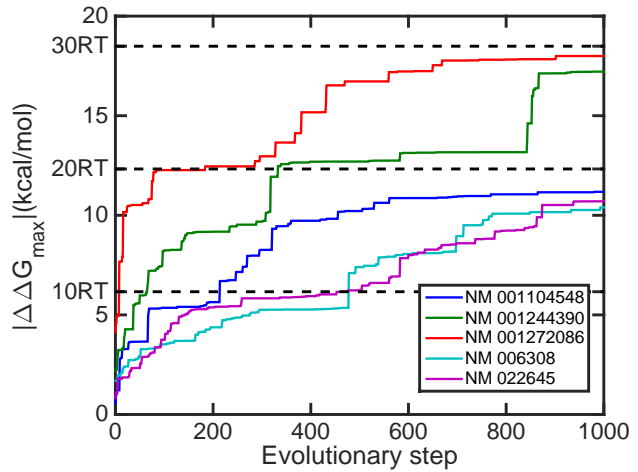}
\caption{Cooperativities along artificial evolutionary trajectories starting from five human sequences seeking to maximize the cooperativity free energy. Compared to the cooperativities in natural human sequences, which are on the order of $RT$, the artificial sequences generated by mutations support much stronger cooperativities of $|\dGmeasure| > 10RT$.}% \texttt{change y axis to $|\dGmeasure_{\mathrm max}|$}}
	\label{fig:RNAmutate}
\end{figure}

In order to find out how strong cooperativities between the 3'UTR and the 5'UTR can possibly become, we do not restrict ourselves to natural sequences but rather search for the maximum of the cooperativity free energy $\dGmeasure$ among a huge number of possible sequences using an evolutionary algorithm (see Materials and Methods). In Fig.~\ref{fig:RNAmutate} we show evolution trajectories starting from five human sequences over 1000 mutation steps. At the end of the artificial evolution trajectories all five human mRNAs mutate to artificial sequences comprising very strong cooperativities. Three sequences NM\_001104548, NM\_006308, and NM\_022645 eventually mutate to sequences with $|\dGmeasure_{\mathrm max}| \approx 20RT$, and NM\_001244390 and NM\_001272086 have even higher final cooperativities of $|\dGmeasure_{\mathrm max}| \approx 30RT$.  This result shows that appropriately designed RNA molecules can support cooperativities much stronger than typically found in natural mRNA sequences. However, considering that most biochemical reactions occur at energy scales of a few $RT$, such large $\dGmeasure$ are actually unrealistic and inappropriate for biological functions as the strong interactions they would confer would not be reversible thus rendering regulation impossible.

\section{DISCUSSION}

In this study we have introduced a quantitative measure of RNA
secondary structure based protein binding cooperativity and we have
demonstrated that significant cooperativity between protein binding
sites at the two ends of an RNA molecule is a generic feature of RNA
secondary structure formation.  We have come to this conclusion by
systematically studying the ensembles of human mRNAs, {\it C. elegans} mRNAs, and random sequences generated from human mRNAs, in which the sequence lengths are between 500nt and
1500nt, and verified the effect in selected longer human mRNAs.

The cooperativity free energy $\dGmeasure$ rises up to a biologically
relevant value (i.e. $|\dGmeasure| > RT \approx 0.6$kcal/mol) in more than 1000 among the 2282 human mRNAs we investigated.  Our statistical
investigation of these sequences shows that this cooperativity gets
stronger as the two binding sites get closer to their respective ends,
suggesting a significant ``end-to-end'' cooperativity between two
proteins or microRNAs that bind on the opposing ends of an mRNA
molecule.  We interpret this observed cooperativity to imply that the
two binding partners, albeit far from each other {\em when measured in
  terms of nucleotides along the mRNA sequence}, can actually be close
to each other {\em in physical and secondary structure space} due to
the intricate secondary structures of the mRNA molecule. In
consequence, the binding of the two partners becomes interdependent,
and this effect allows for combinatorical post-transcriptional
regulation.

We compared the cooperativity free energies $\dGmeasure(\Dout)$ of human mRNAs and two ensembles of the random sequences generated from the human mRNAs. We discovered that all three sequence ensembles support biologically relevant cooperativity, $|\dGmeasure| > RT$ and have maximal cooperativities when the binding sites are relatively close to the ends of the UTRs. However, in human mRNAs the cooperativity $\dGmeasure$ decreases rapidly as the distance $\Dout$ increases; on the other hand, the cooperativity $\dGmeasure(\Dout)$ of random sequences decays smoothly, as shown in Fig.~\ref{fig:STDcompare}. We view the difference as an evidence that human mRNAs have distinguishable cooperativity properties compared to random sequences and note that the fact that human sequences appear to show a weaker cooperativity than their random counterparts can be a consequence of us only looking at symmetrically located putative binding sites that likely do not coincide with the true pairs of binding sites in human UTRs.  Consistently with this, when we investigated experimentally annotated microRNA binding sites, we did find a group of microRNA binding sites with a strong cooperativity free energy around $\dGmeasure = -3.8{\rm kcal/mol} \approx -6RT$ that is statistically overrepresented compared to randomly permuted sites.

An intriguing phenomenon about RNA secondary structure induced
cooperativity is its symmetry in terms of the sign of the
cooperativity.  The symmetry of the distribution of cooperativity free
energies shown in Fig.~\ref{fig:dGhistComp_2D} implies that the number
of mRNAs with positive cooperativity is roughly equal to the number of
those with negative cooperativity. Whether this symmetry is a generic
property of RNA or if it is specific to the ensembles of mRNAs studied here,
is an interesting topic for future investigations.

Another interesting fact we find about the cooperativity between the two ends of an RNA is that it appears exclusively determined by the 5' and 3'UTR sequences. Our investigation of {\it C. elegans} mRNAs shows a positive relationship between UTR length and cooperativity strength. Moreover, an ensemble of sequences generated by substituting a fixed CDS for the CDSs of human mRNAs shows very similar cooperativities to the ensemble of original human mRNAs. These two results support the picture that the 5' and 3'UTRs ``communicate" with each other with little interference from the CDS.

In our study we first chose the protein binding sites to be symmetrically
arranged, i.e., at equal distances from their respective ends of the
mRNA molecule.  Based on the properties of RNA secondary structure, we
would expect that {\em statistically} asymmetric configurations of the
protein binding site will behave the same as the symmetric
configurations with the outside distance $\Dout$ being the only
relevant parameter.  However, when studying the cooperativity free
energy of a {\em specific} molecule, being able to vary the positions
of both binding sites independently should dramatically increase the
search space and thus enable even higher cooperativity free energies.
Unfortunately, the computational cost of varying the two positions
independently is prohibitive 
%\st{ for all but short sequences at this point}
since each combination of binding sites requires a new folding of the
entire molecule.

However, we were still able to investigate asymmetric configurations in some limited cases. In our study of microRNAs, the 3'UTR binding sites are fixed, and thus we were able to tune the 5'UTR binding sites and search through a range of different $n_1$ with respect to the asymmetric, fixed $n_2$. Also, when we looked for the maxima of cooperativity in artificially evolved RNA sequences, we constrained the calculation of $\dGmeasure$ in a small range of asymmetric $n_1$ and $n_2$, as a compromise given the limitations imposed by computational cost.  In this case we found very significant cooperativity free energies up to $|\dGmeasure_{\mathrm{max}}|\approx30RT$.

An important aspect for future investigations is the inclusion of
sequence specificity.  For the purpose of this work, we assumed
perfect sequence specificity and allowed the protein (or
  microRNA) to only bind at one pre-defined location along the mRNA.
With techniques like RNAcompete~\cite{Ray09,Ray13} systematically measuring
the affinities of many important RNA binding proteins to {\em all}
possible binding sites, it should be possible to allow sequence
dependent binding of two proteins to be incorporated into RNA
secondary structure prediction and we will pursue this avenue in order
to identify mRNAs and pairs of proteins that actually employ this
mechanism of post-transcriptional regulation and could be verified
experimentally.  Lastly, while our study entirely focused on
thermodynamical quantities, studying the kinetics of the interplay of
RNA structure and protein binding will be important as well.

\section{ACKNOWLEDGEMENTS}

We thank Nikolaus Rajewsky and Michael Poirier for fruitful discussions initiating this work and the anonymous reviewers for suggesting interesting additional directions incorporated into this work.

\section{FUNDING}

National Science Foundation under grants No. DMR-01105458 and DMR-1410172.

\subsubsection{Conflict of interest statement.} None declared.
%\newpage

\end{document}